\documentclass[rnote]{aa}
\usepackage{graphicx}
\usepackage{natbib}
\bibpunct{(}{)}{;}{a}{}{,}
\input epsf.sty

\def \grss {GRS 1915+105~}
\def \grs {GRS 1915+105}

\def \sax {{\it Beppo}SAX }

\begin{document}

\title{Time properties of the  the $\rho$ -class burst of the microquasar GRS~1915+105 observed 
with {\it Beppo}SAX in April 1999. }
\author{T.~Mineo\inst{1}
\and F.~Massa\inst{2,}\thanks{retired}
\and E.~Massaro\inst{3,4}
\and A.~D'A\`{i}\inst{1}}
\institute{INAF, IASF Palermo,via U. La Malfa 153, I-90146 Palermo, Italy
\and INFN, Sezione Roma 1, Piazzale A. Moro 2, I-00185 Roma, Italy 
\and INAF, IAPS, via del Fosso del Cavaliere 100, I-00113 Roma, Italy
\and In Unam Sapientiam, Piazzale A. Moro 2, I-00185 Roma, Italy 
}
\offprints{mineo@ifc.inaf.it}
\date{Received ....; accepted ....}
\markboth{F.~Massa et al.: Timing properties of the microquasar GRS~1915+105 in the $\rho$ class}
{F.~Massa et al.: Timing properties of the microquasar GRS~1915+105 in the $\rho$ class}

\abstract{We present a temporal analysis of a \sax  observation of \grss performed on April 13, 1999 when the  source was in the $\rho$  class, which is  characterised by quasi-regular bursting activity. 
The aim of the present work is to confirm and extend the validity of  the results obtained with
a \sax observation performed on October 2000 on the recurrence time of the burst and on 
the hard X-ray delay. 
We divided the entire data set into several series, each corresponding to a satellite
orbit, and performed the Fourier and wavelet analysis and the limit cycle mapping technique using  
the count rate and the average energy as independent variables. 
We found that the count rates correlate  with the recurrence time of bursts 
and with hard X-ray delay, confirming the results previously obtained.
In this observation,    however,   the recurrence times are distributed along two parallel branches
with a constant difference of 5.2$\pm$0.5 s. }


\authorrunning{T.~Mineo et al.}
\titlerunning{Timing properties of the microquasar GRS~1915+105 in the $\rho$ class}

\maketitle

\section{Introduction}

The bursting behaviour \citep[see the review paper by][]{Fender2004} characterises the $\rho$ class of \grss X-ray emission \citep{Belloni2000}. It  was discovered in 1996 with Rossi-XTE \citep{Taam1997} and interpreted as the result of instabilities in an accretion disk, as expected by some previous theoretical calculations \citep{Taam1984}.
This particular behaviour of \grs,  known in the literature  as {\it \textup{heartbeat}}, exhibits the properties of  a limit cycle \citep[e.g.][]{Szuszkiewicz1998, Janiuk2005} that has been  extensively studied 
under many  aspects \citep[see, for instance,][]{Neilsen2011, Neilsen2012}.
The discovery of  a similar phenomenon in  IGR J17091$-$3624  \citep{Altamirano2011}
and MXB 1730-335 \citep[Rapid Burster,][]{Bagnoli2015}  
makes the interest in these studies more general, and comparative analyses can be useful 
to obtain a more complete picture of the physics  underlying these cycles.

In  some previous papers \citep[][hereafter Papers I,  II, and III, respectively]{Massaro2010, Mineo2012, Massa2013},  we investigated the properties of the burst series emitted  by the micro-quasar \grss in the $\rho$ class. 
In particular, using the data of a long \sax observation performed in October 2000, 
we reported several interesting correlations between the mean burst recurrence time 
$T_{rec}$ and the X-ray photon count rate.
We developed a method for measuring the hard X-ray delay ({\it HXD}) 
based on the loop trajectories described in the  count rate -- mean photon energy ($CR$-$E$) plane 
during the bursts (Paper III). We found that HXD, which basically
is the time separation between the midpoints of the count rate and mean photon energy  curves of each burst (see Sect.~\ref{HXD}), is  related to the brightness level of the  source.
The aim of the present analysis is to confirm and extend the validity of our previous results with
different independent data and to explore to which extent these correlations are stable.
We considered another observation of  {\it Beppo}SAX, performed in April 1999, whose analysis has never been published. 
 We applied the same methods as used in the previous studies,
which we do not describe here for the sake of brevity.

\section{Observation and data reduction}
\label{observation}

We analysed the \sax  observation of \grss performed on April 13, 1999. 
We used data obtained with the Medium Energy Concentrator Spectrometer (MECS)  in the 1.6--10~keV energy band \citep{Boella1997}, which
were retrieved from the {\it Beppo}SAX archive at the ASI Science Data Center (ASDC).
Data were reduced and selected following the standard procedure and using the SAXDAS 
v. 2.3.3 package. MECS events were selected within a circle of 8$'$ radius that contains about 95\% of the point source signal. 
No background  was considered because the source  rate is two orders of magnitude higher than the contemporary background level evaluated in a region far from the source. 
Its subtraction is then irrelevant in the analysis.

We divided the entire data set into several series, each corresponding to a satellite
orbit, and named them according to the same criterion as adopted in Paper I for the 2000 data.
Each series corresponds to a continuous observing period and is tagged with the 
letter P followed by the orbit sequential number; if a particular series is interrupted because of telemetry gaps, a letter is added to the series' numbers.  
The analysis is performed over segments with exposures longer than 1000 s to have sufficient statistics. Table~\ref{log99} lists the codes of the considered 18 series,  the starting times, the exposure lengths,  the number of bursts, and other quantities derived from our analysis.

\begin{table*}
\caption{Observation log of the 1999 observation: ObsId 20985001.
The columns report the labels of the time series, the starting times of each series from 
T$_0$=13 April 1999 2:11:34 UT, the exposures, the average 
rates, and the parameters derived from our analysis. } 
\label{log99}
\begin{center}
\begin{tabular}{lrccccccc}
\hline\hline
 Series  &  Tstart (s)  & Exposure & \multicolumn{2}{c}{Average MECS rate } & N bursts & $T_{rec}$ & {\it HXD} & Type$^*$ \\
         &   \multicolumn{1}{c}{s}    & \multicolumn{1}{c}{s}  &   burst   & $BL$  &    &  s & s & \\
\hline
    P1    &  0.0       & 2417.6  &  232.4 & 147.1 & 48 & 51.1 & 6.3 & S \\
    P2b   &  6215.3    & 2011.5  &  226.0 & 141.6 & 41 & 50.0 & 5.9 & S \\ 
    P3b   &  12443.3   & 1600.1  &  211.0 & 132.5 & 35 & 46.9 & 5.5 & S \\ 
    P4a   &  16654.2   & 1135.2  &  207.2 & 132.5 & 25 & 46.8 & 5.1 & S \\
    P4b   &  18666.3   & 1019.9  &  214.9 & 134.1 & 22 & 47.8 & 5.5 & S \\                    
    P5a   &  22484.0   & 1436.5  &  205.8 & 129.2 & 32 & 45.7 & 5.1 & T \\
    P6a   &  28254.5   & 1785.1  &  209.0 & 127.1 & 43 & 42.3 & 3.9 & S \\
    P7    &  34019.3   & 2162.4  &  209.9 & 125.2 & 53 & 41.3 & 3.7 & S \\
    P8    &  39785.6   & 2496.2  &  203.7 & 117.9 & 60 & 42.1 & 3.3 & T \\
    P9    &  51343.3   & 3104.6  &  206.3 & 122.3 & 73 & 43.0 & 3.7 & S \\
    P10   &  56973.5   & 3255.2  &  198.3 & 115.9 & 78 & 42.1 & 3.3 & S \\
    P11   &  62746.5   & 3246.8  &  197.6 & 118.5 & 77 & 42.5 & 3.8 & S \\                     
    P12a  &  68722.5   & 1530.2  &  198.7 & 119.4 & 57 & 44.4 & 4.0 & S \\
    P12b  &  69926.5   & 1914.0  &  198.7 & 119.4 & 57 & 44.5 & 4.0 & S \\
    P13   &  74448.7   & 3127.2  &  196.6 & 119.9 & 71 & 44.5 & 4.2 & S \\
    P14a  &  80451.5   & 1150.2  &  200.3 & 122.5 & 26 & 45.7 & 4.3 & S \\                 
    P15   &  86617.5   & 2467.3  &  204.2 & 125.5 & 53 & 47.3 & 4.9 & S \\
    P16b  &  92842.4   & 1587.3  &  207.3 & 132.1 & 25 & 51.9 & 5.3 & S \\
\hline 
\end{tabular}
\end{center} 
$^*$ This parameter is related to the Fourier power spectrum of the series: S only one peak is present, T two peaks are detected.
\end{table*}

\section{Data analysis and results of the 1999 $\rho$ -class observation}
\label{romode}

The temporal analysis of data in $\rho$ mode,  performed taking into account the results 
presented in Papers I and III, was focused on evaluating the recurrence 
time of the bursts and of the HXD. 

\begin{figure}[ht]
\vspace{0.2 cm}
\centerline{
\includegraphics[width=8.0cm,angle=0]{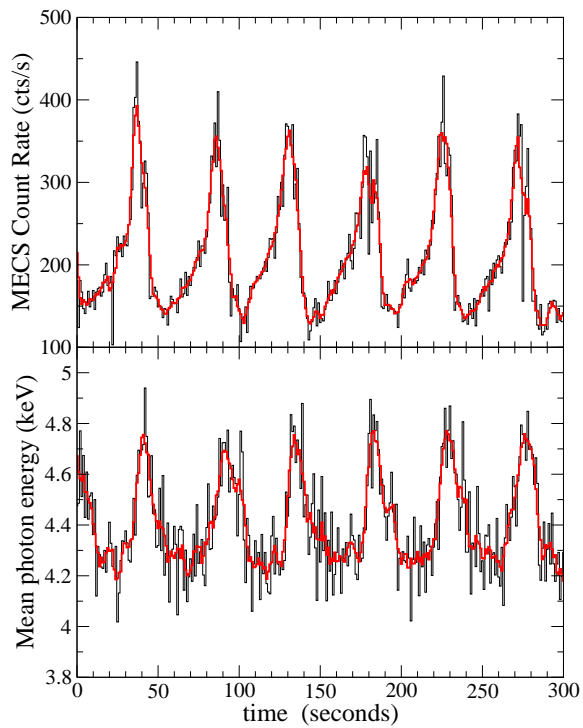}
}
\caption{Time evolution of the series P3b.
The top panel shows a short portion of the light curve in the 
entire MECS energy range; the bottom panel presents the relative mean energy.
Black curves are the rates with 1 s integration time, while red curves are averaged
in a 5 s interval.}
\label{curvesP3}
\end{figure}

\begin{figure}[ht]
\hspace{-0.5 cm}
\centerline{
\includegraphics[width=6.6cm,angle=-90]{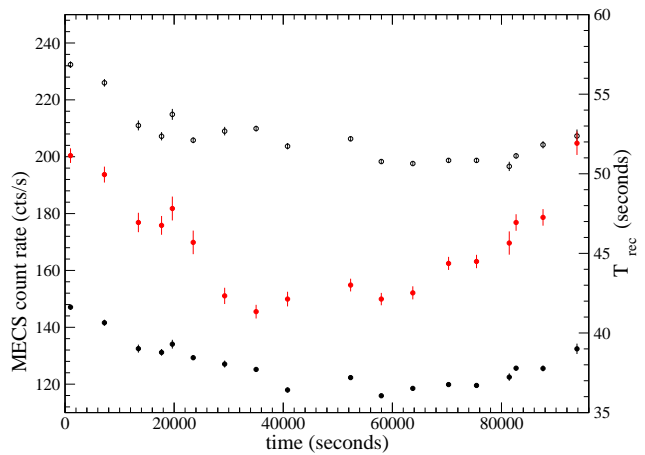}
}
\caption{Time evolution  (left scale)  of the average burst (open black circle) and mean $BL$ (black circle) count rate  and  of  $T_{rec}$ of bursts (red points, right scale) during the observation in April 1999. 
}
\label{timeCRTrec}
\end{figure}

\subsection{Recurrence time of bursts}

All  data series present regular sequences of bursts typical of the $\rho$ class, 
characterised by a slow leading trail ({\it SLT}), a {\it \textup{pulse,}} and a final decaying 
trail ({\it FDT}) (see Paper I) superposed onto a baseline level ($BL$) that remains 
remarkably constant in each series. The $BL$  rate was evaluated in the three seconds after the minimum between two consecutive bursts.
A 300 s long segment of the series P3b is shown in the upper panel of Fig.~\ref{curvesP3}.

\begin{figure}[ht]
\vspace{0.2 cm}
%\centerline{
\includegraphics[width=6.7cm,angle=-90]{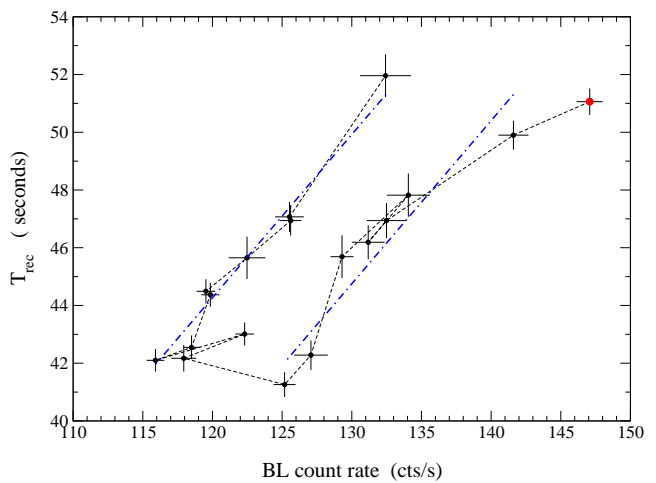} 
\caption{T$_{rec}$ vs the $BL$ rate during the April 1999
observation.
The dot-dashed blue lines are the linear best fits to the data of the two
branches with the same slope.
The initial point is marked by a red  circle.
The dotted lines connecting the points describe the time sequence of data. 
}
\label{DcrTrec}
\end{figure}

A small but significant variation of the total brightness with a relative amplitude of 
$\sim$18\%, from a mean count rate of 232.4 ct/s at the beginning of the observation 
to 196.6 ct/s in the middle, is clearly visible (see Fig.~\ref{timeCRTrec}).  
A similar trend is also detected in the $BL$ light curve, but with higher  amplitude  
($\sim$27\%).
We note that the dynamical range of  the observed variability remains within the range 
detected in October 2000 (see Paper I). 

As in Paper I, we evaluated the recurrence time of bursts by means of Fourier power spectra
of each series. These periodograms generally presented a single sharp dominant peak
(S type, according to the definition given in \citet{Massaro2010}) with the exception of only two series that presented two peaks (T type) with a quite small separation (see Table~\ref{log99}).
The values of $T_{rec}$,  also given in Table~\ref{log99},  follow 
the same trend as the count rate, as shown in Fig.~\ref{timeCRTrec}, in agreement 
with our findings in the October 2000 observation. We note, however, that its variation ($\sim$26\%)
has an amplitude that is almost identical to those of  the $BL$ rate. 

The plot in  Fig.~\ref{DcrTrec} shows the relation between 
$T_{rec}$ and the $BL$ count rate  starting from the first series marked by a red point. 
It is interesting to note that  points are located in two clearly separated 
branches: the lower branch corresponds to the initial decreasing rate time
interval, and the upper branch occurred during the subsequent increasing
source count rate. 
No effect like this was detected in the observation of October 2000 because the count 
rate never decreased (Paper I).

To quantify the systematic difference in  $T_{rec}$ between the two branches, we computed the linear fits to the  data  (excluding the first point) and obtained similar  values for the slopes;
 we then fixed both values to the mean value.  The constant difference
between the two lines (see  Fig.~\ref{DcrTrec})  is 5.2$\pm$0.5 s, about 11\% of the mean $T_{rec}$ in the pointing.

\begin{figure}[ht]
\vspace{0.2 cm}
\centerline{
\includegraphics[width=6.7cm,angle=-90]{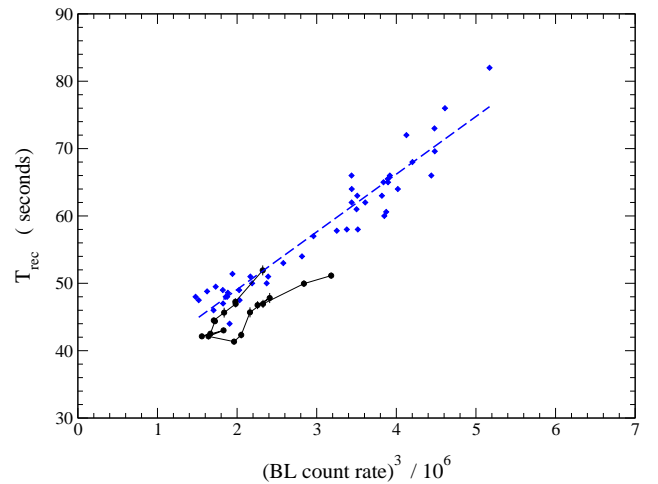}}
\caption{Comparison of the $T_{rec}$ vs $BL$ count rate between
observations made in April 1999 
(black filled circles) and October 2000 (blue diamonds).}
\label{T_rec}
\end{figure}

We found that $T_{rec}$ is proportional to the third power of the $BL$ count rate, as shown in Fig.~\ref{T_rec}.
For comparison, the values relative to the October 2000 observation 
are also plotted in the same figure: the upper branch of the 1999 data agrees well with the 
2000 data, while the decreasing branch lies systematically below these points.
When we fitted all data (1999 and 2000) with a power law, we obtained an exponent equal to 
3.1$\pm$0.1.

\subsection{Mapping in the {\it CR-E} plane and the HXD}
\label{HXD}
\begin{figure}[ht]
\centerline{
\vbox{
\includegraphics[width=7.0cm,angle=-90]{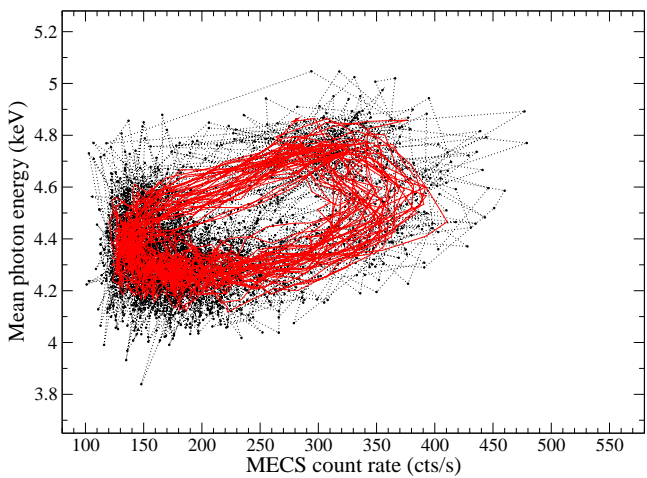}
}}
\caption{Trajectory in the  CR-E plane for 1 s integration time (black curve) and 5 s integration time (red curve).
}
\label{loopCRE_P3}
\end{figure}

\begin{figure}[ht]
%\hspace{0.5 cm}
\centerline{
\includegraphics[width=8.0cm,angle=0]{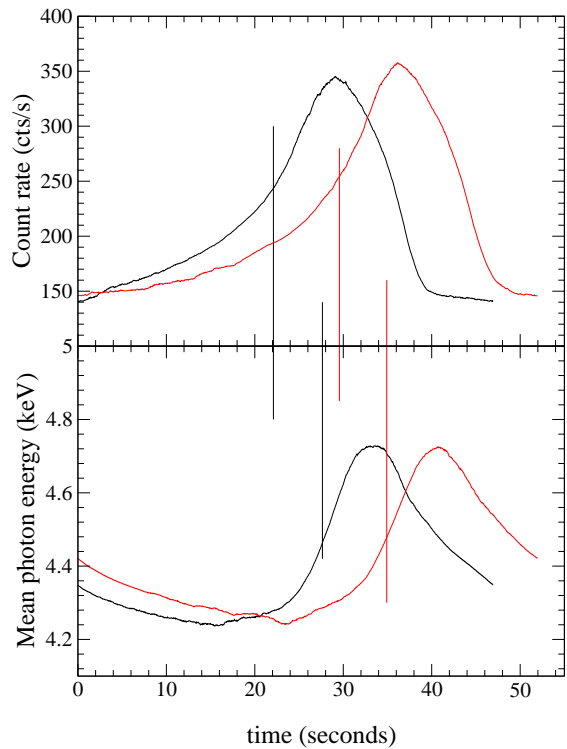}
}
\caption{Comparison between the average burst profiles (top panel) in the series P3b 
(black) and P16b (red) and between the relative mean energies (bottom panel).
Vertical lines show how {\it HXD} is measured from the midpoints times of the two data
series.} 
\label{meanpulsesP3P16}
\end{figure}

Applying the same approach as described in Paper III, we computed 
the data series of the mean photon energy (see the lower panel in Fig.~\ref{curvesP3}), 
which were used to construct the curves in the {\it CR-E} plane.
Again we found that each series describes a well-defined loop of approximately 
elliptical shape.
An example that is useful for illustrating the limit cycle typical 
of the $\rho$ class  is given in Fig.~\ref{loopCRE_P3} for the P3b data; the other series 
exhibit very similar trajectories.
The structure of the limit cycle is related to a delay between the count rate 
(photon luminosity) and the mean photon energy (temperature) curves.
This delay of high-energy photons is clearly apparent from the mean burst 
shape and energy curve plotted in Fig.~\ref{meanpulsesP3P16} for the two data 
series P3b (black curves) and P16 (red curves), which  correspond to points in the 
decreasing and increasing branch that have  similar $BL$  rates but average $T_{rec}$ equal to 46.0 s and 51.9 s, respectively.
We computed these mean curves  by means of the algorithm applied to the {\it CR-E}
trajectories; this is explained in detail in \citet{Massaro2014}.

\begin{figure}[ht]
\vspace{0.2 cm}
\includegraphics[width=6.7cm,angle=-90]{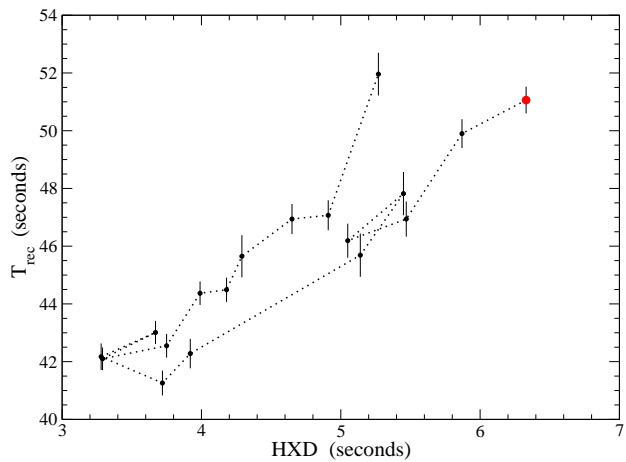}
\caption{ T$_{rec}$ vs {\it HXD} the  during the April 1999 observation.
The initial point is marked by the red  circle.
The dotted lines connecting the points describe the temporal evolution of data. 
}
\label{HXDTrec}
\end{figure}

The definition of {\it HXD} used in our works (see Paper III) is graphically explained  in Fig.~\ref{meanpulsesP3P16} with the vertical lines: it measures
 the time separation  between the corresponding midpoints of the two curves for each bursts.
The resulting mean values {\it HXD} for each series are also given in Table~\ref{log99}: 
similar as for the 2000 data, a positive correlation between the delay and $T_{rec}$ was 
found, as shown in Fig.~\ref{HXDTrec}, where a two-branch path is also clearly apparent.

In Paper III we found that HXD is proportional to the third power of the $BL$ count rate;
a relation similar to that of  $T_{rec}$  .  
We verified that the correlation between {\it HXD} and the cube of the $BL$ 
rates  also holds for these data.

\section{Summary and conclusion}
\label{conclusion}

\begin{figure}[ht]
\vspace{0.2 cm}
\includegraphics[width=6.7cm,angle=-90]{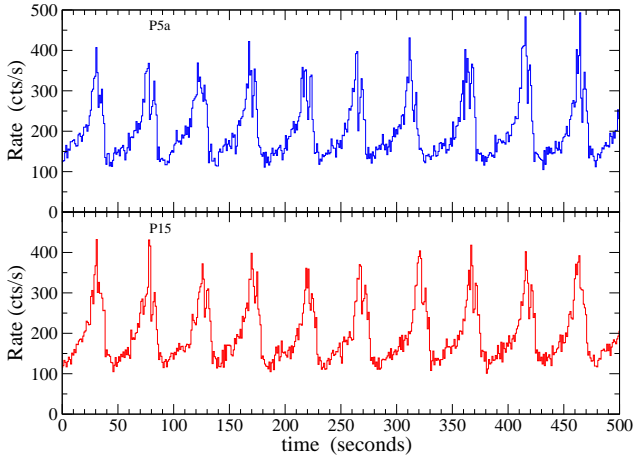}
\caption{Segments of the 1-6--10 keV series P5a (top panel) and P15 (bottom panel)
in the decreasing and increasing branch, respectively.
No change of the burst structure is apparent.}
\label{P5a_vs_P15}
\end{figure}

We analysed a \sax observation of  \grss performed in April 1999 when it was in the $\rho$ -variability class and compared the results with those of another previously studied longer pointing of October 2000.
The main properties of the $\rho$ class were confirmed: the positive correlation between the recurrence time of bursts and the $BL$ rate, the closed trajectories in the $CR$-$E$ plane, and the occurrence of an {\it HXD} that was also correlated with both $T_{rec}$ and rate. 
In this observation, however,  the recurrence times $T_{rec}$ are distributed along two parallel branches with a constant difference  of 5.2$\pm$0.5 s.
The peaks' structure does not show any significant change between the two branches, as clearly apparent in Fig.~\ref{P5a_vs_P15}, where two segments of the series P5a and P15 are plotted.

A physical explanation for this double branch effect is unclear, as is the mechanism producing the class bursting.   \citet{Taam1997}  proposed for the first time that the bursting originates from a non-linear thermal-viscous instability in the accretion disk. However, a  dynamical approach 
to reproduce the observed general structures has not been developed so far. 
In a first attempt,  \citet{Massaro2014} applied the well-known FitzHugh-Nagumo dynamical
equations in two non-dimensional variables $x$ and $y$, which was found to be proportional to the photon flux and disk temperature, respectively.
Their time evolution is given by two differential equations, one of which contains a 'cooling' term in $x^3$, which is necessary to achieve an unstable equilibrium point around which a limit cycle can be established:

\begin{equation}
\left\{
\begin{array}{lcl}
dx/dt &=& - \rho x^3 + \chi x - \gamma y -  J  \\
dy/dt &=& x - y  
\end{array}\right.
.\end{equation}

These equations have four parameters, one of which is a forcing $J$ that \citet{Massaro2014}
 proposed to be related to the mass accretion rate in the disk.
They demonstrated that changes of its value affect both the recurrence time of the bursts and the mean level of intensity.
Moreover, a change of the same parameter can also move the system to a stable equilibrium point with a transition to a non-bursting behaviour, for example, as observed in the $\chi$ class
\citep{Belloni2000}.
The present finding of the two branches of the $T_{rec}$-$BL$ rate plot shown in Fig.~\ref{HXDTrec} suggests that in addition to $J,$  at least one of the other structural parameters, for instance $\chi$ or $\gamma,$ should be considered as variable in the burst modelling.
A further development of this mathematical model and the physical interpretation of these changes is beyond the goals of this article, however.

\begin{acknowledgements}
The authors thank the anonymous referee for improving the paper with constructive comments and suggestions. They also thank the personnel of ASI Science Data Center,  particularly M. Capalbi, for help in retrieving \sax archive data. 
\end{acknowledgements}

\bibliographystyle{aa}
\bibliography{grs1915_1}

\begin{thebibliography}{15}
\expandafter\ifx\csname natexlab\endcsname\relax\def\natexlab#1{#1}\fi

\bibitem[{{Altamirano} {et~al.}(2011){Altamirano}, {Belloni}, {Linares}, {van
  der Klis}, {Wijnands}, {Curran}, {Kalamkar}, {Stiele}, {Motta},
  {Mu{\~n}oz-Darias}, {Casella}, \& {Krimm}}]{Altamirano2011}
{Altamirano}, D., {Belloni}, T., {Linares}, M., {et~al.} 2011, \apjl, 742, L17

\bibitem[{{Bagnoli} \& {in't Zand}(2015)}]{Bagnoli2015}
{Bagnoli}, T. \& {in't Zand}, J.~J.~M. 2015, \mnras, 450, L52

\bibitem[{{Belloni} {et~al.}(2000){Belloni}, {Klein-Wolt}, {M{\'e}ndez}, {van
  der Klis}, \& {van Paradijs}}]{Belloni2000}
{Belloni}, T., {Klein-Wolt}, M., {M{\'e}ndez}, M., {van der Klis}, M., \& {van
  Paradijs}, J. 2000, A\&A, 355, 271

\bibitem[{{Boella} {et~al.}(1997){Boella}, {Chiappetti}, {Conti}, {Cusumano},
  {del Sordo}, {La Rosa}, {Maccarone}, {Mineo}, {Molendi}, {Re}, {Sacco}, \&
  {Tripiciano}}]{Boella1997}
{Boella}, G., {Chiappetti}, L., {Conti}, G., {et~al.} 1997, \aaps, 122, 327

\bibitem[{{Fender} \& {Belloni}(2004)}]{Fender2004}
{Fender}, R. \& {Belloni}, T. 2004, \araa, 42, 317

\bibitem[{{Janiuk} \& {Czerny}(2005)}]{Janiuk2005}
{Janiuk}, A. \& {Czerny}, B. 2005, \mnras, 356, 205

\bibitem[{{Massa} {et~al.}(2013){Massa}, {Massaro}, {Mineo}, {D'A{\`i}},
  {Feroci}, {Casella}, \& {Belloni}}]{Massa2013}
{Massa}, F., {Massaro}, E., {Mineo}, T., {et~al.} 2013, \aap, 556, A84

\bibitem[{{Massaro} {et~al.}(2014){Massaro}, {Ardito}, {Ricciardi}, {Massa},
  {Mineo}, \& {D'A{\`i}}}]{Massaro2014}
{Massaro}, E., {Ardito}, A., {Ricciardi}, P., {et~al.} 2014, \apss, 352, 699

\bibitem[{{Massaro} {et~al.}(2010){Massaro}, {Ventura}, {Massa}, {Feroci},
  {Mineo}, {Cusumano}, {Casella}, \& {Belloni}}]{Massaro2010}
{Massaro}, E., {Ventura}, G., {Massa}, F., {et~al.} 2010, \aap, 513, A21+

\bibitem[{{Mineo} {et~al.}(2012){Mineo}, {Massaro}, {D'Ai}, {Massa}, {Feroci},
  {Ventura}, {Casella}, {Ferrigno}, \& {Belloni}}]{Mineo2012}
{Mineo}, T., {Massaro}, E., {D'Ai}, A., {et~al.} 2012, \aap, 537, A18

\bibitem[{{Neilsen} {et~al.}(2011){Neilsen}, {Remillard}, \&
  {Lee}}]{Neilsen2011}
{Neilsen}, J., {Remillard}, R.~A., \& {Lee}, J.~C. 2011, \apj, 737, 69

\bibitem[{{Neilsen} {et~al.}(2012){Neilsen}, {Remillard}, \&
  {Lee}}]{Neilsen2012}
{Neilsen}, J., {Remillard}, R.~A., \& {Lee}, J.~C. 2012, \apj, 750, 71

\bibitem[{{Szuszkiewicz} \& {Miller}(1998)}]{Szuszkiewicz1998}
{Szuszkiewicz}, E. \& {Miller}, J.~C. 1998, \mnras, 298, 888

\bibitem[{{Taam} {et~al.}(1997){Taam}, {Chen}, \& {Swank}}]{Taam1997}
{Taam}, R.~E., {Chen}, X., \& {Swank}, J.~H. 1997, \apjl, 485, L83+

\bibitem[{{Taam} \& {Lin}(1984)}]{Taam1984}
{Taam}, R.~E. \& {Lin}, D.~N.~C. 1984, \apj, 287, 761

\end{thebibliography}

\end{document}